\documentclass[journal]{IEEEtran}

\usepackage{cite}
\usepackage{amsmath,amssymb,amsfonts}
\usepackage{algorithmic}
\usepackage{graphicx}
\usepackage[caption=false]{subfig}
\usepackage{textcomp}
\usepackage{xcolor}
\usepackage{color, soul}
\usepackage{enumitem}
\usepackage{import}
\usepackage[ruled,vlined]{algorithm2e}
\usepackage{setspace}
\usepackage{hyperref}
\usepackage{tabularx}
\usepackage{xcolor}
\usepackage{hyperref}
\usepackage{comment}
\usepackage{multirow}

\ifCLASSINFOpdf
\else
\fi

\begin{document}

\title{Faster than Real-Time Simulation:\\ Methods, Tools, and Applications}

\author{\IEEEauthorblockN{\textbf{XiaoRui Liu, Juan Ospina, Ioannis Zografopoulos, Alonzo Russell,  Charalambos Konstantinou}\\}
\IEEEauthorblockA{Department of Electrical and Computer Engineering, FAMU-FSU College of Engineering \\
Center for Advanced Power Systems, Florida State University\\
Email:\{xliu9, jjospina, izografopoulos, aarussell,   ckonstantinou\}@fsu.edu}
\vspace{-8mm}}


\maketitle
\pagenumbering{gobble}

\begin{abstract}
Real-time simulation enables the understanding of system operating conditions by evaluating simulation models of physical components running synchronized at the real-time wall clock. Leveraging the real-time measurements of comprehensive system models, faster than real-time (FTRT) simulation allows the evaluation of system architectures at speeds faster than real-time. FTRT  simulation can assist in predicting the system's behavior efficiently, thus assisting the operation of system processes. Namely, the provided acceleration can be used for improving system scheduling, assessing system vulnerabilities, and predicting system disruptions in real-time systems. The acceleration of simulation times can be achieved by utilizing digital real-time simulators (RTS) and high-performance computing (HPC) architectures. FTRT simulation has been widely used, among others, for the operation, design, and investigation of power system events, building emergency management plans, wildfire prediction, etc. In this paper, we review the existing literature on FTRT simulation and its applications in different disciplines, with a particular focus on power systems. We present existing system modeling approaches, simulation tools and computing frameworks, and stress the importance of FTRT accuracy.

\end{abstract}


\IEEEpeerreviewmaketitle

\vspace{-3mm}
\section{Introduction}

Real-time simulation refers to the simulation of system models operating at the same rate as the actual ``wall clock'' time. It employs a fixed discrete time-step and requires that the system states can be solved in this predetermined time-step, otherwise \emph{overruns} will occur (i.e., leading to non-real-time solutions) \cite{rtsimulation}. 
Real-time simulation facilitates the synchronous interconnection -- leveraging the wall clock time as the ground truth --  of physical devices (e.g., controllers) with the simulation system models. As a result, the performance of hardware devices can be evaluated in a realistic environment and under real-time constraints,  de-risking the device migration into the actual system. For example, real-time simulation is extensively used as part of design and planning tools for electric power systems (EPS) as well as for testing grid equipment before field deployment. 
Different from real-time, faster than real-time (FTRT) simulation can be used to predict the behavior of ``ultra-large systems" by utilizing the real-time system measurements and operational attributes.  FTRT simulation requires less time than the time-step used by real-time simulation, allowing multiple FTRT simulations to be performed between two consecutive real-time simulation time-steps. Therefore, FTRT can be utilized to predict the propagation of events (e.g., cascading instability failures, cyberattacks, out-of-step (OOS) conditions for generators, etc.) and assist in corrective decision-making processes. 

\begin{figure*}[t]
\centering
    \subfloat[]{
        \includegraphics[width=0.6\textwidth]{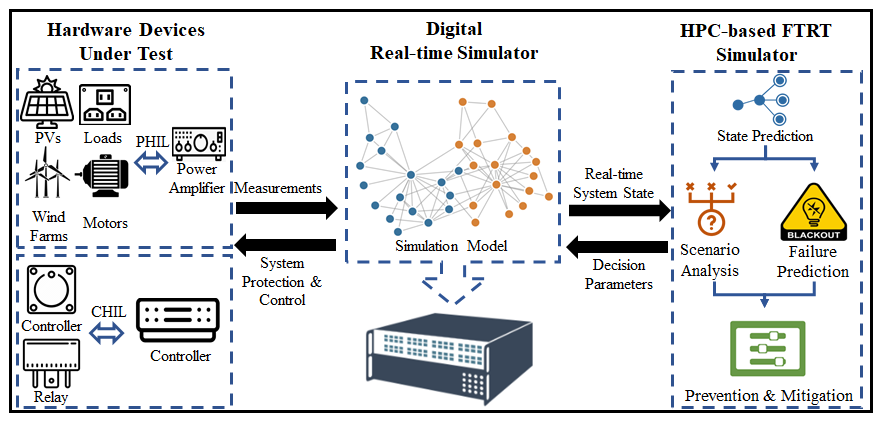}
        \label{fig:architecture}
    } 
    \subfloat[]{
        \includegraphics[width=0.35\linewidth]{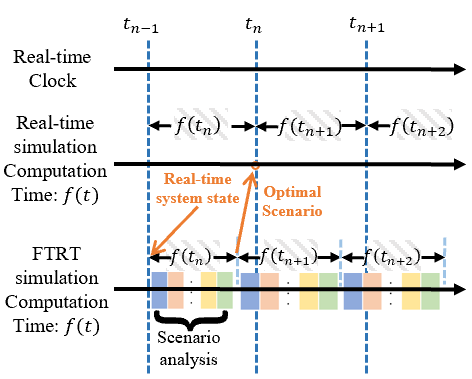}
        \label{fig:timings}
    } \\
    
\caption[CR]{\subref{fig:architecture} An overview flowchart of the interaction between the hardware devices with a HPC-based FTRT simulator, \subref{fig:timings} FTRT simulation timing details.} 
\vspace{-5mm}
\label{fig:FTRT}
\end{figure*}

In order to achieve FTRT simulation, both the integration of a digital real-time simulator (RTS) and a high-performance computing (HPC)-based simulator is essential. As shown in Fig. \ref{fig:architecture}, the measurement data of the hardware devices (e.g., power hardware-in-the-loop (PHIL) and controller hardware-in-the-loop (CHIL)) are collected and sent to the RTS through a communication interface. The RTS is used to compute the next system state based on the power system model and the current system states (e.g., voltage/current at buses, equipment operation, etc.). The real-time state information is transferred to an HPC-based simulator through a high-speed real-time communication interface. The HPC-based simulator is capable of processing data according to different computational demands. For example, in applications such as fault mitigation, the HPC-based environment can predict potential power system state trajectories leveraging the real-time system states (provided by the RTS), thus avoiding cascading failures or equipment compromises. The computational resources of HPC-based simulators expedite the evaluation of system equations, realizing significantly faster execution times when compared to traditional real-time simulation (Fig. \ref{fig:timings}). 




The power grid resilience can be enhanced when FTRT simulation is employed to analyze the dynamic time-varying behavior of EPS and its transient response to events (e.g., faults). 
FTRT simulations can predict potential cascading failures and preemptively issue protection and remediation actions.
 For example, 
the authors in \cite{8854219}, outline the effectiveness of FTRT simulation in predicting the system dynamics of an integrated AC/DC grid and define effective control schemes to mitigate unforeseen contingencies that could disturb the grid stability. 
Furthermore, FTRT allows system operators to simulate the physical system models under attack, identify potential impacts, and provide prevention or mitigation plans. For instance, the work in \cite{8592818}, replicates fault events using a FTRT simulation setup to study the dynamic post-fault behavior of fault-induced dynamic voltage recovery (FIDVR) events. 
The proposed FTRT digital representation can detect the FIDVR scenarios and predict their impact on the system, enabling system operators to thwart such events and prevent them before they escalate to system-wide disruptions.

Although FTRT simulation can assist in estimating the system operating conditions, challenges still exist when applied to different simulation scenarios.  
{These include difficulties in \textit{(i)} system modeling, \textit{(ii)} enhancement of computational efficiency, and \textit{(iii)} accuracy improvement. The trade-off for FTRT system modeling is directly connected with the fidelity and complexity of the differential equations that represent each component in the simulated system. Such components include generators, controllers, and induction motor models, which contain nonlinear elements and become computationally intensive to solve at a FTRT rate. The complex system modeling and the parallel computation of multiple scenarios bring additional hurdles impacting the overall execution time, while the accuracy and fidelity of such models should not be sacrificed when accelerating simulations.} In this work, we present a survey of FTRT simulation-based studies. The main contributions of this paper are listed below:

\begin{itemize}
    \item We provide an overview of FTRT simulation studies and their applications for power systems and other disciplines.  
    \item We focus on the current developments and methodologies of FTRT simulation for power system applications. 
    \item 
    {We describe FTRT system modeling methods, simulation environments, and discuss accuracy requirements. }
\end{itemize}

The rest of the paper is organized as follows. Section \ref{s:Related} presents the related work and Section \ref{s:challenges} introduces the existing FTRT simulation approaches. Finally, Section \ref{s:conclusion} concludes the paper and provides directions for future work.
\vspace{-2mm}
\section{Related work} \label{s:Related}

Significant efforts have been exerted in FTRT simulation studies with the objective of improving system operation in critical infrastructures 
as well as enhancing emergency response plans (e.g., during wildfires, natural disasters, etc.). In this section, we outline the main topics of existing FTRT simulation literature for power system related applications as well as in other disciplines. 
{A summary of the reviewed papers, based on their research topic, system size, and simulation time-step is presented in Table \ref{table:review}.}  
\vspace{-3mm}

\begin{table*}[t]
\small
\centering
\caption{
{Research topic categories and example literature for FTRT simulation.}}
\label{table:review}
{
\begin{tabular}{||c|c|c|c|c|c||}
\hline\hline

\multicolumn{2}{||c|}{\textbf{Category}} &\textbf{Example Literature}&\textbf{System} &\textbf{Time-step}& \textbf{Ref.} \\\hline

 & & Parallel computing of DAEs &14,000
-bus&-&\cite{6495836}\\\cline{3-6}
 
 &\multirow{2}{*}{Computational} &\multirow{2}{*}{Solving the nonlinear DAEs on reconfigurable parallel hardware}& x2 39-bus &\multirow{2}{*}{200µs} &\multirow{2}{*}{\cite{8854219}}\\ 
 
 & &&+ HVDC&&\\\cline{3-6}
 
&\multirow{2}{*}{challenges} & Customized parallel hardware architecture tailored to DAEs&14-bus&5µs&\cite{6381508}\\ \cline{3-6}

&&PGNME based method for solving large set of DAEs &45552-bus&1ms&\cite {sullivan2017faster} \\\cline{3-6}
&&Adaptive time-stepping based scheme for ensuring FTRT &39-bus&5-20µs&\cite{8818631}\\ \cline{2-6}

Power & Prediction   &\multirow{2}{*}{Cyberattacks intrusion detection} &Subsystem of & &\multirow{2}{*}{\cite{7741974}}\\ 
system & of&&Greenland grid&3ms&\\ \cline{3-6}

related&system&Short-term voltage instability prediction &9-bus&1ms& \cite{8592818}\\ \cline{3-6}

& behaviors&Generator OOS prediction&39-bus&50µs& \cite{abedini2017generator}\\\cline{2-6}

&  & Eliminate the short term impact of the FIVDR& 39-bus &1ms&\cite{8571803} \\ \cline{3-6}
&Mitigation&\multirow{2}{*}{Reduce the impact of SSI in a power grid with wind penetration}&11-bus&\multirow{2}{*}{200µs}&\multirow{2}{*}{\cite{9055075}}\\ 
&schemes&&+HVDC&&\\ \hline

& Emergency & Fast
fluid dynamics method based airflow studies in
buildings&Building&0.02-1s& \cite{zuo2009}\\ \cline{3-6}

&management& Emergency navigation for providing the evacuation route &Shopping mall&-&\cite{7134084} \\ \cline{2-6}

& Prediction& \multirow{2}{*}{Wildfire propagation modeling and predicting} &$1500m^{2}$&\multirow{2}{*}{10s}&\multirow{2}{*}{\cite{grasso2020}} \\

Others& /detection&&area&& \\ \cline{3-6}

&{system behavior}& Acoustic events detection &Room \cite{carletta2007unleashing}&-& \cite{lin2012improving} \\ \cline{2-6}

&{System}&{Facial alignment}& -&-& {\cite{bhagavatula2017faster}}\\ \cline{3-6}

&improvements&Drinking-water-treatment-plant simulation&-&-&\cite{worm2012training}\\

\hline \hline
\end{tabular}}
\vspace{-5mm}
\end{table*}

\vspace{-1mm}
\subsection{Power System Related Applications of FTRT Simulations}

Due to the complexity of power systems, it is challenging to simultaneously solve sets of nonlinear ordinary differential equations (ODEs) (e.g., required for electromagnetic transient (EMT) simulations), or differential-algebraic equations (DAEs) (e.g., necessary for transient stability (TS) simulations), for large-scale EPS without exceeding the real-time time-step. To improve the computational performance of simulations, a parallel computing mechanism is examined in \cite{6495836}. The HPC-based model is designed to improve the execution time by implementing a parallelized computing process for the models of the EPS under test. 
The proposed FTRT implementation reaches a solution 3 times faster than a conventional real-time simulation. 
Namely, the benchmark used is the Western Electricity Coordinating Council (WECC) power system that is composed of about $4,555$ buses and $3,000$ generators supplying electricity to $71$ million users. Similarly, a fine-grained relaxation algorithm (FGRA) is demonstrated in \cite{8854219}, which can solve the nonlinear DAEs using parallelized and pipelined hardware. 
According to the authors, the hardware emulation of the two case studies is accelerated by $134$ times, 
{with a total $7,462$ clock-cycle hardware latency.}

A reconfigurable hardware-based simulator is developed in \cite{6381508} to achieve EPS simulation in FTRT rates. The proposed method adaptively matches the hardware architecture of the computation engine with the power system structure representing the mathematical system model. The computation engine enforces parallelism at the sub-systems/components, equations, and primitive operations level. HPC can support the parallel general Norton multiport equivalent (PGNME)-based domain decomposition method, enhancing the performance of FTRT TS simulation for large-scale systems \cite {sullivan2017faster}. The PGNME technique decomposes the power system network into sub-systems that can be evaluated in parallel. 
 The simulation results of the approximated WECC achieve speeds 10 times faster than the benchmark. Furthermore, a variable time-stepping scheme is utilized to accelerate 
FTRT in \cite{8818631}. In this scheme, the time-step of each sub-system is different, reducing the overall processing latency. However, all the sub-systems need to be synchronized (regardless of their respective time-step). Therefore, the time-step of each sub-system needs to be linearly proportional to the global simulation time-step. For example, to achieve synchronization between sub-systems with longer time-steps (e.g., generator models, lighting transients), sub-systems with shorter time-steps (e.g., only transmission lines) might be computed several times. 



Besides the improvement of computing efficiency, the prediction of the system behavior is another advantage of FTRT simulations. From the cybersecurity perspective, the authors in \cite{7741974} introduce an intrusion detection scheme for maliciously altered data based on a distributed dynamic state estimator. The proposed detection scheme is achieved through a command authentication method that leverages a FTRT implementation. 
Thus, the prediction and protection against vulnerabilities are ensured by determining the security impact that specific commands can have on the system model. Another prediction-based example is presented in \cite{8592818}, where the authors predict FIDVR events that will occur in a short-term horizon (less than $30s$) via FTRT dynamic simulations. The proposed algorithm utilizes the collected data from phasor measurement units (PMUs) and monitors the topological changes of the system for potential faults that could trigger FIVDR events (e.g., line outage faults, three-phase short circuit faults, etc.). Once an event is detected, the prediction algorithm is activated to assess the post-fault behavior. Leveraging FTRT for the prediction of OOS condition of turbine generator units
is demonstrated in \cite{abedini2017generator}. An OOS condition occurs when one of the generators experiences a large angular difference, which could result in severe oscillations in the voltage, current, and torque measured at the system level. This study shows that the proposed mechanism can detect the OOS $5.86s$ faster than a traditional real-time simulation-based implementation for the New England IEEE 39-bus system. 
\vspace{-3mm}



Many research works focus on FTRT simulation for preventing threats targeting power systems. For example, the authors in \cite{8571803} propose mitigation strategies to circumvent short-term voltage instabilities caused by FIVDR. Namely, an under-voltage load shedding scheme is utilized to respond to FIVDR events and maintain system stability. The minimum amount of load that needs to be shed 
is predicted leveraging FTRT simulations and a digital replica of the system. 
Researchers in \cite{9055075} present a vulnerability named subsynchronous interaction (SSI) introduced by capacitive series compensation when boosting the capacity of transmission lines in long-distance power delivery. 
SSI can result in severe overvoltage events and trip the associated transmission facilities \cite{NERCSSI}. 
In order to negate the impacts of SSI, such as immediate oscillations in the shaft torque, the authors present a method to mitigate the influence of SSI in a hybrid AC/DC grid. Notably, the grid topology of this case study supports wind farm integration and utilizes FTRT dynamic simulations to predict crucial control parameters (e.g., active/reactive power flows and injections). 
{The reported hardware computational overhead occupies $4,174$ clock cycles.} 


\vspace{-4mm}
\subsection{Applications of FTRT Simulations in Other Disciplines }

The benefits of FTRT simulations can assist other disciplines by ensuring awareness and enhancing the system operation. Typical applications include building emergency management strategies, prediction/detection of natural emergencies, and improving the performance of specific system mechanisms. For example, the authors in \cite{zuo2009} develop a fast fluid dynamics method to compute the airflow and temperature distributions in buildings. 
 The proposed method can predict the airflow patterns $4$ to $100$ times faster than real-time simulation. 
This, in turn, can be used to enhance the deployment of emergency management systems designed to predict fire, smoke, or contamination 
in the airflow of buildings. Moreover, in \cite{7134084}, the authors show that FTRT simulations can be used to model emergency navigation systems designed to provide `optimal' evacuation routes and effectively increase human survival rates during unexpected disasters. A fire-related scenario is used to examine the efficacy of the proposed building evacuation algorithm. 
When a person discovers the fire, the current fire location can be identified by taking snapshots from the actual areas inside the building. The photos will then be transferred to image processing servers and activate the emergency navigation system. The fire spreading model is simulated on a cloud-based system that predicts the fire spreading path. Based on the initial distribution of evacuees, the optimal routes can be identified 
using FTRT simulations.

FTRT simulations have also been employed in wildfire propagation studies, forecasting both the size and the shape of the burnt area during natural disasters \cite{grasso2020}. Although the computational overhead increases with the system size, FTRT simulation can predict the fire dynamics
in $58s$ for every minute of the real-time wall clock. Another application of FTRT simulation is reported in \cite{lin2012improving}, where a saliency-maximized audio spectrogram is proposed to enable audio browsing and detect acoustic events $10$ times faster than real-time simulations. Similarly, for image processing, FTRT simulation is used for 3D facial alignment, where models are capable of landmarking approximately $170$ faces per second \cite{bhagavatula2017faster}. FTRT simulation has also been employed in the acceleration of training of experienced or inexperienced personnel (e.g., employees in a water plant or laymen) using the drinking water treatment plant simulation models presented in \cite{worm2012training}. 
\vspace{-2mm}
\section{FTRT Simulation Methods} \label{s:challenges}

Different approaches are currently used to 
expedite FTRT simulation implementations. The overall performance of FTRT simulation heavily depends on the discretization methods used to simulate the system and the parallelization methods used to exploit software and hardware architectures. In this section, we present: \textit{(i)} the modeling approaches used in real-time simulation and FTRT simulation implementations in both power system-related and other disciplines' applications,  \textit{(ii)} FTRT simulation platforms and tools utilized in power system studies, and \textit{(iii)} the accuracy requirements for evaluating FTRT simulation results of the power system applications. 

\vspace{-5mm}

\subsection{Modeling Approaches for Real-Time and FTRT Simulations} 

Models represent mathematical equations and/or data that are used to explain and predict the behavior of systems. For complex systems, such as EPS, both nonlinear ODEs and DAEs are used to represent them. 
A general formulation of such system equations is shown in Eq. (\ref{Eq:DEAs}),  where $\mathbf{x}$ represents the state variables vector (e.g., generator angle, generator speed), $\mathbf{y}$ is the vector of variables (e.g., bus voltage, bus angle, or load flow variables), $\mathbf{f}$ represents a derivative function of the state variable $\mathbf{x}$ (e.g., the dynamic behavior of generator flux linkages, rotor angles, control states, and dynamic load states \cite{singh2001direct}),  and $\mathbf{g}$ denotes a function of the outputs. In this subsection, we present 4 modeling approaches leveraged by FTRT implementations.  
{Although the mathematical formulation for FTRT algorithms does not differ from conventional real-time simulation modeling, different implementations are utilized by FTRT methods to leverage parallel architectures and enhance computation efficiency.}


\begin{equation}
\begin{array}{l}
\mathbf{\dot{x}(t)}=\mathbf{f}(\mathbf{x(t)}, \mathbf{y(t)}) \\
\mathbf{g}(\mathbf{x(t)}, \mathbf{y(t)})=0
\label{Eq:DEAs}
\end{array}
\end{equation}

\noindent\subsubsection{Discretization Methods}

\begin{itemize}[leftmargin=*, wide=0pt, label={}]
    \item \underline{Mathematical Formulation}: Models need to be discretized before their simulation in digital environments regardless of the simulation type used (e.g., offline or real-time). 
    The discretization is performed using accurate numerical discretization methods that transform continuous-time differential equations into discrete difference equations. Some of the most commonly used numerical discretization methods are the forward and backward Euler-method, the forward difference method, and the Runge-Kutta methods. 
    Other discretization methods, known as Z-transforms, can also be used to discretize continuous-time system models aiding convergence. 
     Some of the most common Z-transform methods are the step-invariant transformation (SIT), the ramp-invariant transformation (RIT), and the time-shifted SIT (TSSIT) methods \cite{gurrala}.

    \item \underline{FTRT approach}: The increased complexity of higher-order differential equations requires more iterations for solvers to reach accurate solutions. To address these issues, the authors in \cite{9055075} utilize a $4$th order Runge-Kutta model to solve high-order ($>9$th) DAEs. To achieve FTRT execution times, researchers in \cite{8854219} propose a parallel execution of the system model that introduces a processor parallelism method for the $9$-machine power system. Each machine is fully independent (i.e., each machine only connects to one bus), and thus, the results can be calculated in a distributed manner.

\end{itemize}

\noindent \subsubsection{Predictor-Corrector Method}

\begin{itemize}[leftmargin=*, wide=0pt, label={}]
    \item \underline{Mathematical Formulation}: The \textit{Predictor-Corrector} method presented in \cite{8274505} consists of 2 modules, i.e., the \textit{predictor} and the \textit{corrector}. The value $\mathbf{x}_{n}^{pred}$ in the \textit{predictor} is calculated based on the current value $\mathbf{x}_{n-1}$ of the system as shown in Eq. (\ref{Eq:predictor}). On the other hand, the \textit{corrector} is in charge of updating the initial predicted value to the corrected value ($\mathbf{x}_{n}^{corr}$) via Eq. (\ref{Eq:corrector}). The term $h$ represents the model step-size. Both processes can run in parallel for FTRT applications. 
    
    \vspace{-1mm}

    \begin{equation}
    \mathbf{x}_{n}^{pred}=\mathbf{x}_{n-1}+h f\left({t}_{n-1},x_{n-1}\right)
    \label{Eq:predictor}
    \end{equation}
    
    \vspace{-6mm}
    
    \begin{equation}
    \mathbf{x}_{n}^{corr}=\mathbf{x}_{n-1}+\frac{h}{2}(f\left({t}_{n-1},x_{n-1}\right)+f\left({t}_{n},x_{n}^{pred}\right))
    \label{Eq:corrector}
    \end{equation}
    
    \item \underline{FTRT approach}: In order to achieve FTRT simulation execution times, parallel computing schemes are employed. A prominent example that leverages the concept of parallelism for FTRT simulations based on a \textit{Predictor-Corrector} method is presented in \cite{8274505}. The evaluation of these two processes is performed in parallel and on different processors to speed up the overall process. This simulation has been performed in the GridPACK HPC framework (introduced in Section \ref{s:Simulation Environment}) with detailed dynamic models containing generators, controllers, and nonlinear loads.

\end{itemize}

\noindent \subsubsection{Linear Multi-step Method}

\begin{itemize}[leftmargin=*, wide=0pt, label={}]
    \item \underline{Mathematical Formulation}: This approach uses multiple linear step-sizes for the system simulations. 
     The \textit{linear multi-step} approach proposed in \cite{8818631} starts from an initial value and updates the new state values using subsequent multi-step sizes. This method is mathematically formulated as shown in Eq. (\ref{Eq:LMM}), where coefficients $a$ and $b$ are determined by the users, and $s$ denotes the total number of steps. The coefficients $a$ and $b$ require balancing since they directly affect the approximation of the true solution and the simulation speed. The term $h$ represents a constant step-size. 
    
    \vspace{-1mm}
    
    \begin{equation}
    \sum_{j=0}^{s} a_{j} x_{n+j}=h \sum_{j=0}^{s} b_{j} f\left(t_{n+j}, x_{n+j}\right)
    \label{Eq:LMM}
    \end{equation}

    \item \underline{FTRT approach}: The authors in \cite{8818631} utilize an adaptive multi-time-step approach (i.e., $h$ can change for different sub-systems) to effectively improve the processing latency of the FTRT implementation. Hardware coordination between different sub-systems is required to accelerate the simulation leveraging FPGA-based emulation. Details about the sub-system coordination are given in Section \ref{s:Simulation Environment}. 
    
\end{itemize}

\noindent \subsubsection{Reconfigurable-Hardware Decoupling Method}

\begin{itemize}[leftmargin=*, wide=0pt, label={}]
    \item \underline{Mathematical Formulation}:  Hardware processor parallelism methods can also be exploited for the segmentation and simulation of large-scale EPS.
    Following this approach, the authors in \cite{6381508} directly decouple the EPS model into sub-systems which can be evaluated in parallel.  
    Every system state variable can be calculated using $V_{i}(j)=\sum_{k=1}^{n} Y_{i}^{-1}(j, k) I_{i}(k)$, where $Y_i$, $V_i$, and $I_i$ are the admittance matrix, node voltages, and current values of each sub-system, respectively. Consequently, the solution can be reached through matrix-vector multiplication. 
    Since each product is calculated independently on a dedicated FPGA, $n^2$ processing devices are required to compute $Y_{i}^{-1}$ with its corresponding $I_{i}$ for this sub-system with $n$ nodes.

    \item \underline{FTRT approach}: In \cite{6381508}, a reconfigurable-hardware real-time power system simulator is presented that can be adapted for FTRT purposes. The main idea behind the proposed real-time simulation is to enable the parallel execution of large-scale EPS models by running different subparts of the EPS network on FPGAs. This solution provides speed, scalability, and reconfigurability, allowing the creation of `modifiable' clusters adapted for real-time and FTRT simulation purposes. Another approach exploiting hardware reconfiguration processor parallelism is the decomposition method proposed in \cite{sullivan2017faster}. The decomposition scheme is based on multiport equivalence theory designed to significantly decrease simulation time. 
    {In \cite{8854219}, the proposed reconfigurable parallel architecture solves non-linear AC/DC system equations using the iterative Newton-Raphson method. System decomposition improves the computation time and hardware resource utilization by correlating state $x_i$ with the $i_{th}$ row of the Jacobian matrix.}
    
    \end{itemize}

\vspace{-5mm}

\subsection{FTRT Simulation Frameworks and Tools}\label{s:Simulation Environment}

Due to the complexity of modeling large-scale dynamic networks, computational burden reduction is critical to ensure that the overall simulation processing can be achieved within the real-time interval. Evaluating system solutions sequentially or using single-processor computer systems might be insufficient to achieve FTRT timescales. For example, the authors in \cite{8274505} report that it takes around $60s$ to run a $20s$ simulation for a WECC system composed by a detailed generator and controller model on a computer running PowerWorld with an Intel Core i7, 2.8 GHz CPU, and 8GB of RAM memory. The lack of tools that can solve complex system models motivates the development of FTRT simulation platforms. Following, we discuss prominent examples of FTRT-based simulation platforms used for power system studies.

\subsubsection{GridPACK HPC Framework}
The GridPACK HPC is an open-source software framework developed by the Pacific Northwest National Laboratory (PNNL) which enables the modeling and simulation of large-scale EPS using HPC-based FTRT platforms \cite{Gridpack}. Specifically, system component modules, such as induction machines, governors, exciters, relays, and network models are provided as packages. The computational load is shared among multiple processors using the provided packages. Each processor is assigned to solve a designated problem (e.g., one sub-network of the whole system) after the dynamic equations and the network equations are partitioned. This network partitioning strategy reduces the network coupling, requires less communication between processors, and streamlines the overall process.


\subsubsection{HPC-based Simulator} 

To efficiently solve TS models for large-scale EPS, the authors in \cite{sullivan2017faster} introduce an HPC-based simulator developed by Mississippi State University. The ``Shadow II'' HPC cluster has 10 nodes, with each node boasting 512GB RAM and 2 Intel E5-2680 processors. 
A message passing interface (MPI) model based on an MPI/OpenMP (Open Multi-Processing) approach is used for the node communications, while multiple threads are executed in parallel in each computing node. The power system network is decomposed into sub-networks using the partitioning tool hMeTiS, minimizing the intercommunication between nodes and balancing the sub-network. Once the decomposition is finalized, a matrix solver deals with the sub-problems.

\subsubsection{FPGA-based Emulation}

In \cite{6381508}, an FPGA-based FTRT simulator for EMT studies is proposed. The simulator exhibits several advantages over traditional RTS such as: \textit{(i)} high-speed (the computation time ranges between tens to hundreds of nanoseconds), \textit{(ii)} scalability (it can be easily scaled up to large-scale systems), \textit{(iii)} reconfigurability (it is flexible and can adapt to different topological requirements), and \textit{(iv)} physical portability (it requires low-power and low-cost hardware). On the other hand, the simulation models should meet three requirements: \textit{(i)} the system model needs to be decoupled allowing parallel execution, 
\textit{(ii)} the models should be strategically designed 
and avoid constant inter-communication between the models and the FTRT simulator, 
and  \textit{(iii)} the hardware architecture should be tailored to the solution of the mathematical model enhancing execution efficiency.

Due to their benefits, FPGA-based systems have been extensively used in FTRT implementations \cite{8854219, 8818631, 9055075, 9324816}. 
For instance, an integrated AC/DC grid simulation is performed on a parallel FPGA-based architecture in \cite{8854219}. Notably, a Xilinx XCVU9P FPGA board is utilized to model a system integrating 2 New England IEEE 39-bus systems and 4 high-voltage DC (HVDC) grids.
Additionally, the authors in \cite{8818631} provide a comprehensive architecture leveraging FTRT hardware emulation for the IEEE 39-bus system using 2 FPGA VCU118 boards. The test system is divided into 11 sub-systems, while transmission lines are further divided into sub-sub systems achieving efficient calculation of the admittance matrices. 
The authors in \cite{9055075} demonstrate that the AC/DC grid can be divided into three modules, wind farm, 4-terminal HVDC grid, and AC grid, in order to minimize the execution time leveraging different time-steps for each module. Moreover, the tasks within one module can be computed in parallel (e.g., converter control (CNT), converter average value model (AVM),and HVDC network inside the HVDC module).







\vspace{-3mm}
\subsection{Accuracy Requirements for FTRT Simulation}
Accuracy requirements for FTRT simulation consider the deviations between the simulated results, the actual system state, and the dynamic system performance. However, a performance-accuracy trade-off exists, i.e., if detailed system models are utilized, higher accuracy will be obtained but at the expense of longer execution time. However, simplified system models reduce computation time by sacrificing accuracy.
Therefore, the FTRT results need to be validated by comparing them with results obtained from other simulation tools evaluating the same mathematical models.   
For instance, the FTRT simulation results provided using the FTRT GridPACK HPC system could be compared with results reported by PowerWorld (offline) or by an RTS (e.g., Opal-RT). In \cite{8274505}, the generator rotor speed results under fault conditions are similar in both systems. 
In \cite{8854219}, the authors validate their FTRT results with an offline TS simulation tool TSAT (DSAToolsTM) by comparing them to the oscilloscope waveforms. 
The FTRT simulation matches the results from TSAT when AC ground faults are performed. 
The authors in \cite{8818631} and \cite{6381508} compare the results of their proposed FTRT model with simulations performed in PSCAD/EMTDC. 
EMT simulations demonstrate that acceptable accuracy levels can be achieved without sacrificing accuracy.

Furthermore, an error analysis between FTRT results and offline Matlab/Simulink results is performed in \cite{9055075} using $\epsilon= ( (R_{F T R T}-R_{\text {Simulink }}) / R_{\text {Simulink }} ) \times 100 \%$, 
where ${R_{F T R T}}$ and $R_{\text {Simulink}}$ refer to the results reported by the FTRT and the Simulink simulations, respectively. The maximum relative errors are $- 0.97\%$ over 4 contingency simulation results including three-phase faults, fault mitigation, overloads, and mitigation after overloads. Thus FTRT simulation accuracy is verified under diverse system contingency scenarios. 

\vspace{-3mm}
\section{Conclusion} \label{s:conclusion}

This paper presents a review of research works employing FTRT simulation. 
Emphasis is drawn on particular applications where FTRT simulations are leveraged to enhance the overall stability of systems. We report how the synergistic relation between FTRT and real-time simulation can be used to prevent or overcome undesirable events. The FTRT predictive capabilities are utilized to monitor and predict system behavior (e.g., fires, natural disasters, voltage instability, etc.) and assist in issuing emergency management plans and/or threat mitigation and aversion policies. In literature, the advantages of FTRT simulation have been primarily explored for the modeling and simulation of large-scale EPS.
This review focuses on how FTRT can: \textit{(i)} deal with the increased computational overhead of large systems, \textit{(ii)} predict the system performance under faults, natural disasters, or renewable energy penetration in real-time applications, and \textit{(iii)} be used to design mitigation strategies and administer threat management plans. 
Future work will focus on harnessing FTRT simulations to reduce power system uncertainty under anomalous operation (e.g., attacks, faults). Furthermore, the capabilities of FTRT simulation will be explored to minimize the impact of operational disruptions in large-scale EPS with renewable energy sources integration such as wind energy, microgrids, PV systems, etc.

\vspace{-2mm}
\bibliographystyle{IEEEtran}
\bibliography{biblio}

\end{document}